\documentclass[conference,10pt,twocolumn,a4paper]{IEEEtran}

\usepackage{cite}
\usepackage[T1]{fontenc}
\usepackage{graphicx}
\usepackage{amssymb}
\usepackage{amsmath}
\usepackage{paralist}
\usepackage{setspace}
\usepackage{microtype}
\usepackage{url}
\usepackage{balance}
\usepackage[caption=false,font=footnotesize]{subfig}
\usepackage{xcolor}
\usepackage{fancyref}
\usepackage{hyperref}
\usepackage{dsfont}

\IEEEoverridecommandlockouts

\usepackage{amssymb}
\usepackage{amsfonts}
\usepackage{mathrsfs}
\usepackage{xspace}
\usepackage{bm}
\usepackage{upgreek}

\newcommand{\safemath}[2]{\newcommand{#1}{\ensuremath{#2}\xspace}}

\safemath{\bma}{\mathbf{a}}
\safemath{\bmb}{\mathbf{b}}
\safemath{\bmc}{\mathbf{c}}
\safemath{\bmd}{\mathbf{d}}
\safemath{\bme}{\mathbf{e}}
\safemath{\bmf}{\mathbf{f}}
\safemath{\bmg}{\mathbf{g}}
\safemath{\bmh}{\mathbf{h}}
\safemath{\bmi}{\mathbf{i}}
\safemath{\bmj}{\mathbf{j}}
\safemath{\bmk}{\mathbf{k}}
\safemath{\bml}{\mathbf{l}}
\safemath{\bmm}{\mathbf{m}}
\safemath{\bmn}{\mathbf{n}}
\safemath{\bmo}{\mathbf{o}}
\safemath{\bmp}{\mathbf{p}}
\safemath{\bmq}{\mathbf{q}}
\safemath{\bmr}{\mathbf{r}}
\safemath{\bms}{\mathbf{s}}
\safemath{\bmt}{\mathbf{t}}
\safemath{\bmu}{\mathbf{u}}
\safemath{\bmv}{\mathbf{v}}
\safemath{\bmw}{\mathbf{w}}
\safemath{\bmx}{\mathbf{x}}
\safemath{\bmy}{\mathbf{y}}
\safemath{\bmz}{\mathbf{z}}
\safemath{\bmzero}{\mathbf{0}}
\safemath{\bmone}{\mathbf{1}}

\bmdefine{\biad}{a}
\bmdefine{\bibd}{b}
\bmdefine{\bicd}{c}
\bmdefine{\bidd}{d}
\bmdefine{\bied}{e}
\bmdefine{\bifd}{f}
\bmdefine{\bigd}{g}
\bmdefine{\bihd}{h}
\bmdefine{\biid}{i}
\bmdefine{\bijd}{j}
\bmdefine{\bikd}{k}
\bmdefine{\bild}{l}
\bmdefine{\bimd}{m}
\bmdefine{\bind}{n}
\bmdefine{\biod}{o}
\bmdefine{\bipd}{p}
\bmdefine{\biqd}{q}
\bmdefine{\bird}{r}
\bmdefine{\bisd}{s}
\bmdefine{\bitd}{t}
\bmdefine{\biud}{u}
\bmdefine{\bivd}{v}
\bmdefine{\biwd}{w}
\bmdefine{\bixd}{x}
\bmdefine{\biyd}{y}
\bmdefine{\bizd}{z}

\bmdefine{\bixid}{\xi}
\bmdefine{\bilambdad}{\lambda}
\bmdefine{\bimud}{\mu}
\bmdefine{\bithetad}{\theta}
\bmdefine{\biphid}{\phi}
\bmdefine{\bideltad}{\delta}

\safemath{\bmia}{\biad}
\safemath{\bmib}{\bibd}
\safemath{\bmic}{\bicd}
\safemath{\bmid}{\bidd}
\safemath{\bmie}{\bied}
\safemath{\bmif}{\bifd}
\safemath{\bmig}{\bigd}
\safemath{\bmih}{\bihd}
\safemath{\bmii}{\biid}
\safemath{\bmij}{\bijd}
\safemath{\bmik}{\bikd}
\safemath{\bmil}{\bild}
\safemath{\bmim}{\bimd}
\safemath{\bmin}{\bind}
\safemath{\bmio}{\biod}
\safemath{\bmip}{\bipd}
\safemath{\bmiq}{\biqd}
\safemath{\bmir}{\bird}
\safemath{\bmis}{\bisd}
\safemath{\bmit}{\bitd}
\safemath{\bmiu}{\biud}
\safemath{\bmiv}{\bivd}
\safemath{\bmiw}{\biwd}
\safemath{\bmix}{\bixd}
\safemath{\bmiy}{\biyd}
\safemath{\bmiz}{\bizd}

\safemath{\bmxi}{\bixid}
\safemath{\bmlambda}{\bilambdad}
\safemath{\bmmu}{\bimud}
\safemath{\bmtheta}{\bithetad}
\safemath{\bmphi}{\biphid}
\safemath{\bmdelta}{\bideltad}

\safemath{\bA}{\mathbf{A}}
\safemath{\bB}{\mathbf{B}}
\safemath{\bC}{\mathbf{C}}
\safemath{\bD}{\mathbf{D}}
\safemath{\bE}{\mathbf{E}}
\safemath{\bF}{\mathbf{F}}
\safemath{\bG}{\mathbf{G}}
\safemath{\bH}{\mathbf{H}}
\safemath{\bI}{\mathbf{I}}
\safemath{\bJ}{\mathbf{J}}
\safemath{\bK}{\mathbf{K}}
\safemath{\bL}{\mathbf{L}}
\safemath{\bM}{\mathbf{M}}
\safemath{\bN}{\mathbf{N}}
\safemath{\bO}{\mathbf{O}}
\safemath{\bP}{\mathbf{P}}
\safemath{\bQ}{\mathbf{Q}}
\safemath{\bR}{\mathbf{R}}
\safemath{\bS}{\mathbf{S}}
\safemath{\bT}{\mathbf{T}}
\safemath{\bU}{\mathbf{U}}
\safemath{\bV}{\mathbf{V}}
\safemath{\bW}{\mathbf{W}}
\safemath{\bX}{\mathbf{X}}
\safemath{\bY}{\mathbf{Y}}
\safemath{\bZ}{\mathbf{Z}}

\safemath{\bZero}{\mathbf{0}}
\safemath{\bOne}{\mathbf{1}}
\safemath{\bDelta}{\mathbf{\Delta}}
\safemath{\bLambda}{\mathbf{\UpLambda}}
\safemath{\bPhi}{\mathbf{\Upphi}}
\safemath{\bSigma}{\mathbf{\Upsigma}}
\safemath{\bOmega}{\mathbf{\Upomega}}
\safemath{\bTheta}{\mathbf{\Uptheta}}

\bmdefine{\biAd}{A}
\bmdefine{\biBd}{B}
\bmdefine{\biCd}{C}
\bmdefine{\biDd}{D}
\bmdefine{\biEd}{E}
\bmdefine{\biFd}{F}
\bmdefine{\biGd}{G}
\bmdefine{\biHd}{H}
\bmdefine{\biId}{I}
\bmdefine{\biJd}{J}
\bmdefine{\biKd}{K}
\bmdefine{\biLd}{L}
\bmdefine{\biMd}{M}
\bmdefine{\biOd}{N}
\bmdefine{\biPd}{O}
\bmdefine{\biQd}{P}
\bmdefine{\biRd}{R}
\bmdefine{\biSd}{S}
\bmdefine{\biTd}{T}
\bmdefine{\biUd}{U}
\bmdefine{\biVd}{V}
\bmdefine{\biWd}{W}
\bmdefine{\biXd}{X}
\bmdefine{\biYd}{Y}
\bmdefine{\biZd}{Z}

\bmdefine{\biDelta}{\Delta}
\bmdefine{\biLambda}{\Lambda}
\bmdefine{\biPhi}{\Phi}
\bmdefine{\biSigma}{\Sigma}
\bmdefine{\biOmega}{\Omega}
\bmdefine{\biTheta}{\Theta}

\safemath{\bimA}{\biAd}
\safemath{\bimB}{\biBd}
\safemath{\bimC}{\biCd}
\safemath{\bimD}{\biDd}
\safemath{\bimE}{\biEd}
\safemath{\bimF}{\biFd}
\safemath{\bimG}{\biGd}
\safemath{\bimH}{\biHd}
\safemath{\bimI}{\biId}
\safemath{\bimJ}{\biJd}
\safemath{\bimK}{\biKd}
\safemath{\bimL}{\biLd}
\safemath{\bimM}{\biMd}
\safemath{\bimN}{\biNd}
\safemath{\bimO}{\biOd}
\safemath{\bimP}{\biPd}
\safemath{\bimQ}{\biQd}
\safemath{\bimR}{\biRd}
\safemath{\bimS}{\biSd}
\safemath{\bimT}{\biTd}
\safemath{\bimU}{\biUd}
\safemath{\bimV}{\biVd}
\safemath{\bimW}{\biWd}
\safemath{\bimX}{\biXd}
\safemath{\bimY}{\biYd}
\safemath{\bimZ}{\biZd}

\safemath{\bimDelta}{\biDelta}
\safemath{\bimLambda}{\biLambda}
\safemath{\bimPhi}{\biPhi}
\safemath{\bimSigma}{\biSigma}
\safemath{\bimOmega}{\biOmega}
\safemath{\bimTheta}{\biTheta}

\safemath{\setA}{\mathcal{A}}
\safemath{\setB}{\mathcal{B}}
\safemath{\setC}{\mathcal{C}}
\safemath{\setD}{\mathcal{D}}
\safemath{\setE}{\mathcal{E}}
\safemath{\setF}{\mathcal{F}}
\safemath{\setG}{\mathcal{G}}
\safemath{\setH}{\mathcal{H}}
\safemath{\setI}{\mathcal{I}}
\safemath{\setJ}{\mathcal{J}}
\safemath{\setK}{\mathcal{K}}
\safemath{\setL}{\mathcal{L}}
\safemath{\setM}{\mathcal{M}}
\safemath{\setN}{\mathcal{N}}
\safemath{\setO}{\mathcal{O}}
\safemath{\setP}{\mathcal{P}}
\safemath{\setQ}{\mathcal{Q}}
\safemath{\setR}{\mathcal{R}}
\safemath{\setS}{\mathcal{S}}
\safemath{\setT}{\mathcal{T}}
\safemath{\setU}{\mathcal{U}}
\safemath{\setV}{\mathcal{V}}
\safemath{\setW}{\mathcal{W}}
\safemath{\setX}{\mathcal{X}}
\safemath{\setY}{\mathcal{Y}}
\safemath{\setZ}{\mathcal{Z}}
\safemath{\emptySet}{\varnothing}

\safemath{\colA}{\mathscr{A}}
\safemath{\colB}{\mathscr{B}}
\safemath{\colC}{\mathscr{C}}
\safemath{\colD}{\mathscr{D}}
\safemath{\colE}{\mathscr{E}}
\safemath{\colF}{\mathscr{F}}
\safemath{\colG}{\mathscr{G}}
\safemath{\colH}{\mathscr{H}}
\safemath{\colI}{\mathscr{I}}
\safemath{\colJ}{\mathscr{J}}
\safemath{\colK}{\mathscr{K}}
\safemath{\colL}{\mathscr{L}}
\safemath{\colM}{\mathscr{M}}
\safemath{\colN}{\mathscr{N}}
\safemath{\colO}{\mathscr{O}}
\safemath{\colP}{\mathscr{P}}
\safemath{\colQ}{\mathscr{Q}}
\safemath{\colR}{\mathscr{R}}
\safemath{\colS}{\mathscr{S}}
\safemath{\colT}{\mathscr{T}}
\safemath{\colU}{\mathscr{U}}
\safemath{\colV}{\mathscr{V}}
\safemath{\colW}{\mathscr{W}}
\safemath{\colX}{\mathscr{X}}
\safemath{\colY}{\mathscr{Y}}
\safemath{\colZ}{\mathscr{Z}}

\safemath{\opA}{\mathbb{A}}
\safemath{\opB}{\mathbb{B}}
\safemath{\opC}{\mathbb{C}}
\safemath{\opD}{\mathbb{D}}
\safemath{\opE}{\mathbb{E}}
\safemath{\opF}{\mathbb{F}}
\safemath{\opG}{\mathbb{G}}
\safemath{\opH}{\mathbb{H}}
\safemath{\opI}{\mathbb{I}}
\safemath{\opJ}{\mathbb{J}}
\safemath{\opK}{\mathbb{K}}
\safemath{\opL}{\mathbb{L}}
\safemath{\opM}{\mathbb{M}}
\safemath{\opN}{\mathbb{N}}
\safemath{\opO}{\mathbb{O}}
\safemath{\opP}{\mathbb{P}}
\safemath{\opQ}{\mathbb{Q}}
\safemath{\opR}{\mathbb{R}}
\safemath{\opS}{\mathbb{S}}
\safemath{\opT}{\mathbb{T}}
\safemath{\opU}{\mathbb{U}}
\safemath{\opV}{\mathbb{V}}
\safemath{\opW}{\mathbb{W}}
\safemath{\opX}{\mathbb{X}}
\safemath{\opY}{\mathbb{Y}}
\safemath{\opZ}{\mathbb{Z}}
\safemath{\opZero}{\mathbb{O}}
\safemath{\identityop}{\opI}

\safemath{\veca}{\bma}
\safemath{\vecb}{\bmb}
\safemath{\vecc}{\bmc}
\safemath{\vecd}{\bmd}
\safemath{\vece}{\bme}
\safemath{\vecf}{\bmf}
\safemath{\vecg}{\bmg}
\safemath{\vech}{\bmh}
\safemath{\veci}{\bmi}
\safemath{\vecj}{\bmj}
\safemath{\veck}{\bmk}
\safemath{\vecl}{\bml}
\safemath{\vecm}{\bmm}
\safemath{\vecn}{\bmn}
\safemath{\veco}{\bmo}
\safemath{\vecp}{\bmp}
\safemath{\vecq}{\bmq}
\safemath{\vecr}{\bmr}
\safemath{\vecs}{\bms}
\safemath{\vect}{\bmt}
\safemath{\vecu}{\bmu}
\safemath{\vecv}{\bmv}
\safemath{\vecw}{\bmw}
\safemath{\vecx}{\bmx}
\safemath{\vecy}{\bmy}
\safemath{\vecz}{\bmz}

\safemath{\veczero}{\bmzero}
\safemath{\vecone}{\bmone}
\safemath{\vecxi}{\bmxi}
\safemath{\veclambda}{\bmlambda}
\safemath{\vecmu}{\bmmu}
\safemath{\vectheta}{\bmtheta}
\safemath{\vecphi}{\bmphi}
\safemath{\vecdelta}{\bmdelta}

\safemath{\matA}{\bA}
\safemath{\matB}{\bB}
\safemath{\matC}{\bC}
\safemath{\matD}{\bD}
\safemath{\matE}{\bE}
\safemath{\matF}{\bF}
\safemath{\matG}{\bG}
\safemath{\matH}{\bH}
\safemath{\matI}{\bI}
\safemath{\matJ}{\bJ}
\safemath{\matK}{\bK}
\safemath{\matL}{\bL}
\safemath{\matM}{\bM}
\safemath{\matN}{\bN}
\safemath{\matO}{\bO}
\safemath{\matP}{\bP}
\safemath{\matQ}{\bQ}
\safemath{\matR}{\bR}
\safemath{\matS}{\bS}
\safemath{\matT}{\bT}
\safemath{\matU}{\bU}
\safemath{\matV}{\bV}
\safemath{\matW}{\bW}
\safemath{\matX}{\bX}
\safemath{\matY}{\bY}
\safemath{\matZ}{\bZ}
\safemath{\matzero}{\bmzero}

\safemath{\matDelta}{\bDelta}
\safemath{\matLambda}{\bLambda}
\safemath{\matPhi}{\bPhi}
\safemath{\matSigma}{\bSigma}
\safemath{\matOmega}{\bOmega}
\safemath{\matTheta}{\bTheta}

\safemath{\matidentity}{\matI}
\safemath{\matone}{\matO}

\safemath{\rnda}{A}
\safemath{\rndb}{B}
\safemath{\rndc}{C}
\safemath{\rndd}{D}
\safemath{\rnde}{E}
\safemath{\rndf}{F}
\safemath{\rndg}{G}
\safemath{\rndh}{H}
\safemath{\rndi}{I}
\safemath{\rndj}{J}
\safemath{\rndk}{K}
\safemath{\rndl}{L}
\safemath{\rndm}{M}
\safemath{\rndn}{N}
\safemath{\rndo}{O}
\safemath{\rndp}{P}
\safemath{\rndq}{Q}
\safemath{\rndr}{R}
\safemath{\rnds}{S}
\safemath{\rndt}{T}
\safemath{\rndu}{U}
\safemath{\rndv}{V}
\safemath{\rndw}{W}
\safemath{\rndx}{X}
\safemath{\rndy}{Y}
\safemath{\rndz}{Z}

\safemath{\rveca}{\bimA}
\safemath{\rvecb}{\bimB}
\safemath{\rvecc}{\bimC}
\safemath{\rvecd}{\bimD}
\safemath{\rvece}{\bimE}
\safemath{\rvecf}{\bimF}
\safemath{\rvecg}{\bimG}
\safemath{\rvech}{\bimH}
\safemath{\rveci}{\bimI}
\safemath{\rvecj}{\bimJ}
\safemath{\rveck}{\bimK}
\safemath{\rvecl}{\bimL}
\safemath{\rvecm}{\bimM}
\safemath{\rvecn}{\bimN}
\safemath{\rveco}{\bomO}
\safemath{\rvecp}{\bimP}
\safemath{\rvecq}{\bimQ}
\safemath{\rvecr}{\bimR}
\safemath{\rvecs}{\bimS}
\safemath{\rvect}{\bimT}
\safemath{\rvecu}{\bimU}
\safemath{\rvecv}{\bimV}
\safemath{\rvecw}{\bimW}
\safemath{\rvecx}{\bimX}
\safemath{\rvecy}{\bimY}
\safemath{\rvecz}{\bimZ}

\safemath{\rvecxi}{\bmxi}
\safemath{\rveclambda}{\bmlambda}
\safemath{\rvecmu}{\bmmu}
\safemath{\rvectheta}{\bmtheta}
\safemath{\rvecphi}{\bmphi}

\safemath{\rmatA}{\bimA}
\safemath{\rmatB}{\bimB}
\safemath{\rmatC}{\bimC}
\safemath{\rmatD}{\bimD}
\safemath{\rmatE}{\bimE}
\safemath{\rmatF}{\bimF}
\safemath{\rmatG}{\bimG}
\safemath{\rmatH}{\bimH}
\safemath{\rmatI}{\bimI}
\safemath{\rmatJ}{\bimJ}
\safemath{\rmatK}{\bimK}
\safemath{\rmatL}{\bimL}
\safemath{\rmatM}{\bimM}
\safemath{\rmatN}{\bimN}
\safemath{\rmatO}{\bimO}
\safemath{\rmatP}{\bimP}
\safemath{\rmatQ}{\bimQ}
\safemath{\rmatR}{\bimR}
\safemath{\rmatS}{\bimS}
\safemath{\rmatT}{\bimT}
\safemath{\rmatU}{\bimU}
\safemath{\rmatV}{\bimV}
\safemath{\rmatW}{\bimW}
\safemath{\rmatX}{\bimX}
\safemath{\rmatY}{\bimY}
\safemath{\rmatZ}{\bimZ}

\safemath{\rmatDelta}{\bimDelta}
\safemath{\rmatLambda}{\bimLambda}
\safemath{\rmatPhi}{\bimPhi}
\safemath{\rmatSigma}{\bimSigma}
\safemath{\rmatOmega}{\bimOmega}
\safemath{\rmatTheta}{\bimTheta}

\usepackage{amssymb}
\usepackage{amsfonts}
\usepackage{mathrsfs}
\usepackage{xspace}
\usepackage{bm}
\usepackage{fancyref}
\usepackage{textcomp}

\usepackage{multirow}
\usepackage{stmaryrd}

\newenvironment{textbmatrix}{	\setlength{\arraycolsep}{2.5pt}%
								\big[\begin{matrix}}{\end{matrix}\big]%
								\raisebox{0.08ex}{\vphantom{M}}}

\def\be{\begin{equation}}
\def\ee{\end{equation}}
\def\een{\nonumber \end{equation}}
\def\mat{\begin{bmatrix}}
\def\emat{\end{bmatrix}}
\def\btm{\begin{textbmatrix}}
\def\etm{\end{textbmatrix}}

\def\ba#1\ea{\begin{align}#1\end{align}}
\def\bas#1\eas{\begin{align*}#1\end{align*}}
\def\bs#1\es{\begin{split}#1\end{split}} 
\def\bg#1\eg{\begin{gather}#1\end{gather}}
\def\bml#1\eml{\begin{multline}#1\end{multline}}
\def\bi#1\ei{\begin{itemize}#1\end{itemize}}

\newcommand{\lefto}{\mathopen{}\left}

\DeclareMathOperator{\tr}{tr}				
\DeclareMathOperator{\sign}{sgn}			
\DeclareMathOperator{\kron}{\otimes}			
\DeclareMathOperator{\Exop}{\opE}			

\newcommand{\Ex}[2]{\ensuremath{\Exop_{#1}\lefto[#2\right]}} 	

\newcommand{\vecnorm}[1]{\lefto\lVert#1\right\rVert}		

\safemath{\dirac}{\delta}					
\safemath{\krond}{\dirac}					

\safemath{\upto}{\uparrow}
\safemath{\downto}{\downarrow}
\safemath{\iu}{j}							
\safemath{\ev}{\lambda}						
\safemath{\hilseqspace}{l^{2}}				
\newcommand{\banachfunspace}[1]{\setL^{#1}}	
\safemath{\hilfunspace}{\banachfunspace{2}}	

\safemath{\SNR}{\textsf{SNR}} 				
\safemath{\PAR}{\textsf{PAR}} 				
\safemath{\No}{N_0}							
\safemath{\Es}{E_s}							
\safemath{\Eb}{E_b}							
\safemath{\EbNo}{\frac{\Eb}{\No}}
\safemath{\EsNo}{\frac{\Es}{\No}}

\DeclareMathOperator{\CHop}{\ensuremath{\opH}} 
\safemath{\tvir}{\rndh_{\CHop}}				
\safemath{\tvtf}{\rndl_{\CHop}}				
\safemath{\spf}{\rnds_{\CHop}}				
\safemath{\bff}{H_{\CHop}}					

\safemath{\ircf}{r_{h}}						
\safemath{\tftvcf}{r_{s}}					
\safemath{\tfcf}{r_{l}}						
\safemath{\bfcf}{r_{H}}						

\safemath{\tcorr}{c_h}						
\safemath{\scf}{c_{s}}						
\safemath{\tfcorr}{c_{l}}					
\safemath{\fcorr}{c_{H}}						

\safemath{\mi}{I}							
\safemath{\capacity}{C}						

\safemath{\normal}{\mathcal{N}}			
\safemath{\jpg}{\mathcal{CN}}			
\safemath{\mchain}{\leftrightarrow}		

\safemath{\dB}{\,\mathrm{dB}}
\safemath{\dBm}{\,\mathrm{dBm}}
\safemath{\Hz}{\,\mathrm{Hz}}
\safemath{\kHz}{\,\mathrm{kHz}}
\safemath{\MHz}{\,\mathrm{MHz}}
\safemath{\GHz}{\,\mathrm{GHz}}
\safemath{\s}{\,\mathrm{s}}
\safemath{\ms}{\,\mathrm{ms}}
\safemath{\mus}{\,\mathrm{\text{\textmu}s}}
\safemath{\ns}{\,\mathrm{ns}}
\safemath{\ps}{\,\mathrm{ps}}
\safemath{\meter}{\,\mathrm{m}}
\safemath{\mm}{\,\mathrm{mm}}
\safemath{\cm}{\,\mathrm{cm}}
\safemath{\m}{\,\mathrm{m}}
\safemath{\W}{\,\mathrm{W}}
\safemath{\mW}{\, \mathrm{mW}}
\safemath{\J}{\,\mathrm{J}}
\safemath{\K}{\,\mathrm{K}}
\safemath{\bit}{\,\mathrm{bit}}
\safemath{\nat}{\,\mathrm{nat}}

\safemath{\define}{\triangleq}			

\safemath{\equivalent}{\sim}
\safemath{\distas}{\sim}					
\safemath{\sdiff}{\Delta}				

\safemath{\reals}{\mathbb{R}}
\safemath{\positivereals}{\reals_{+}}
\safemath{\integers}{\mathbb{Z}}
\safemath{\posint}{\integers_{+}}
\safemath{\naturals}{\mathbb{N}}
\safemath{\posnaturals}{\naturals_{+}}
\safemath{\complexset}{\mathbb{C}}
\safemath{\rationals}{\mathbb{Q}}

\newcommand*{\fancyrefapplabelprefix}{app}		
\newcommand*{\fancyrefthmlabelprefix}{thm}		
\newcommand*{\fancyreflemlabelprefix}{lem}		
\newcommand*{\fancyrefcorlabelprefix}{cor}		
\newcommand*{\fancyrefdeflabelprefix}{def}		
\newcommand*{\fancyrefproplabelprefix}{prop}	
\newcommand*{\fancyrefobslabelprefix}{obs}		
\newcommand*{\fancyrefalglabelprefix}{alg}		
\newcommand*{\fancyrefasmlabelprefix}{asm}	    

\frefformat{vario}{\fancyrefseclabelprefix}{Section~#1}
\frefformat{vario}{\fancyrefthmlabelprefix}{Theorem~#1}
\frefformat{vario}{\fancyreflemlabelprefix}{Lemma~#1}
\frefformat{vario}{\fancyrefcorlabelprefix}{Corollary~#1}
\frefformat{vario}{\fancyrefdeflabelprefix}{Definition~#1}
\frefformat{vario}{\fancyrefobslabelprefix}{Observation~#1}
\frefformat{vario}{\fancyrefasmlabelprefix}{Assumption~#1}
\frefformat{vario}{\fancyreffiglabelprefix}{Fig.~#1}
\frefformat{vario}{\fancyrefapplabelprefix}{Appendix~#1} 
\frefformat{vario}{\fancyrefproplabelprefix}{Proposition~#1}
\frefformat{vario}{\fancyrefalglabelprefix}{Algorithm~#1}
\frefformat{vario}{\fancyrefeqlabelprefix}{(#1)}

\safemath{\dictab}{[\,\dicta\,\,\dictb\,]}

\safemath{\ysig}{\bmy}
\safemath{\ysighat}{\hat{\ysig}}
\safemath{\ysigdim}{M}
\safemath{\xsig}{\bmx}
\safemath{\xsigdim}{N}
\safemath{\nx}{n_x}
\safemath{\zsig}{\bmz}
\safemath{\zsigdim}{\ysigdim}
\safemath{\rsig}{\bmr}
\safemath{\Adict}{\bA}
\safemath{\Adicttilde}{\widetilde{\Adict}}
\safemath{\Adictdim}{\outputdim\times\xsigdim}
\safemath{\avec}{\bma}
\safemath{\avectilde}{\tilde{\avec}}
\safemath{\Bdict}{\bB}
\safemath{\Bdicttilde}{\widetilde{\Bdict}}
\safemath{\Cdict}{\bC}
\safemath{\cvec}{\bmc}
\safemath{\Ddict}{\bD}
\safemath{\Ddictdim}{\ysigdim\times\xsigdim}
\safemath{\dvec}{\bmd}
\safemath{\Ddicttilde}{\widetilde{\bD}}
\safemath{\Bonb}{\bB}
\safemath{\bvec}{\bmb}
\safemath{\Bonbdim}{\ysigdim\times\ysigdim}
\safemath{\noise}{\bmn}
\safemath{\noisedim}{\ysigim}
\safemath{\err}{\bme}
\safemath{\errdim}{\ysigdim}
\safemath{\errset}{\setE}
\safemath{\nerr}{n_e}
\safemath{\delop}{\bP_\errset}
\safemath{\delopc}{\bP_{{\errset}^c}}

\safemath{\cplxi}{\imath}
\safemath{\cplxj}{\jmath}

\safemath{\dict}{\matD}
\safemath{\inputdim}{N}		
\safemath{\outputdim}{M}		
\safemath{\sparsity}{S}	
\safemath{\inputdimA}{{N_a}}	
\safemath{\inputdimB}{{N_b}}	
\safemath{\elemA}{{n_a}}	
\safemath{\elemB}{{n_b}}	
\safemath{\resA}{\matR_a}	
\safemath{\resB}{\matR_b}	
\safemath{\subD}{\matS} 
\safemath{\subA}{\matS_a} 
\safemath{\subB}{\matS_b} 
\safemath{\dicta}{\matA} 	
\safemath{\dictb}{\matB} 	
\safemath{\hollowS}{H}
\safemath{\hollowA}{H_a}
\safemath{\hollowB}{H_b}
\safemath{\cross}{Z}
\safemath{\coh}{\mu_d}			
\safemath{\coha}{\mu_a}			
\safemath{\cohb}{\mu_b}			
\safemath{\mubs}{\nu}	
\safemath{\cohm}{\mu_m} 
\safemath{\dictset}{\setD}	
\safemath{\dictsetp}{\dictset(\coh,\coha,\cohb)}	
\safemath{\dictsetgen}{\dictset_\text{gen}}
\safemath{\dictsetgenp}{\dictsetgen(\coh)}
\safemath{\dictsetonb}{\dictset_\text{onb}}
\safemath{\dictsetonbp}{\dictsetonb(\coh)}

\safemath{\leftside}{U}
\safemath{\rightsideA}{R_a}
\safemath{\rightsideB}{R_b}

\safemath{\indexS}{\setI_S} 

\safemath{\na}{n_a}			
\safemath{\nb}{n_b}			
\safemath{\coeffa}{p_i}	
\safemath{\coeffb}{q_j}	
\safemath{\seta}{\setP}		
\safemath{\setb}{\setQ}     
\safemath{\setw}{\setW}	
\safemath{\setz}{\setZ}	
\safemath{\cola}{\veca}		
\safemath{\colb}{\vecb}		
\safemath{\cold}{\vecd}		
\safemath{\inputvec}{\vecx} 	
\safemath{\error}{\vece}	
\safemath{\noiseout}{\vecz} 	
\safemath{\inputvecel}{x}
\safemath{\inputveca}{\vecx_a}
\safemath{\inputvecb}{\vecx_b}
\safemath{\outputvec}{\vecy}	
\safemath{\lambdamin}{\lambda_{\mathrm{min}}}

\safemath{\elltwo}{\ell_2}
\safemath{\ellone}{\ell_1}
\safemath{\ellzero}{\ell_0}
\safemath{\ellinf}{\ell_\infty}
\safemath{\ellinftilde}{\ell_{\widetilde\infty}}
\safemath{\licard}{Z(\coh,\coha,\cohb)}
\safemath{\xsol}{\hat{x}}
\safemath{\xbord}{x_b}		
\safemath{\xstat}{x_s}		
\safemath{\xstatLone}{\tilde{x}_s}
\safemath{\order}{\mathcal{O}} 
\safemath{\scales}{\Theta} 
\safemath{\ones}{\mathbf{1}} 
\safemath{\zeroes}{\mathbf{0}} 
\safemath{\thlone}{\kappa(\coh,\cohb)} 
\safemath{\constoneA}{\delta} 
\safemath{\constoneB}{\epsilon} 
\safemath{\nlarge}{L}				   
\safemath{\sumlarge}{S_\nlarge}
\safemath{\maxlarger}{P_\nlarge}	   
\safemath{\Pzero}{\textrm{P0}}	
\safemath{\Pone}{\textrm{P1}}
\safemath{\vecfir}{\vecw}			 
\safemath{\vecsec}{\vecz}
\safemath{\elvecfir}{w}              
\safemath{\elvecsec}{z}				 
\safemath{\nlargefir}{n}
\safemath{\normout}{\gamma}
\safemath{\auxfun}{h}
\safemath{\supp}{\textrm{supp}}

\safemath{\indexa}{\ell}
\safemath{\indexb}{r}
\safemath{\indexc}{i}
\safemath{\indexd}{j}

\safemath{\project}{P}

\newcommand{\OBQP}{{\text{QP}}}
\newcommand{\SDR}{{\text{SDR-QP}}}
\newcommand{\LINF}{{\ell_\infty^2\!\text{-QP}}}

\newcommand{\precode}{\mathcal{P}}

\newcommand{\snr}{\rho} 

\begin{document}

%

\title{Nonlinear 1-Bit  Precoding for Massive MU-MIMO with Higher-Order Modulation} 
\author{
\IEEEauthorblockN{Sven Jacobsson$^\text{1,2}$, Giuseppe Durisi$^\text{2}$, Mikael Coldrey$^\text{1}$, Tom Goldstein$^\text{3}$, and Christoph Studer$^\text{4}$} \\ \vspace{-0.1cm}          
\IEEEauthorblockA{$^\text{1}$Ericsson Research, Gothenburg, Sweden; e-mail: \{sven.jacobsson,\,mikael.coldrey\}@ericsson.com}
\IEEEauthorblockA{$^\text{2}$Chalmers University of Technology, Gothenburg, Sweden; e-mail: durisi@chalmers.se}
\IEEEauthorblockA{$^\text{3}$University of Maryland, College Park, MD; e-mail: tomg@cs.umd.edu}
\IEEEauthorblockA{$^\text{4}$Cornell University, Ithaca, NY; e-mail: studer@cornell.edu} 
\vspace{-0.1cm}
\thanks{
SJ and GD were supported by the Swedish Foundation for Strategic Research under grant ID14-0022, and by the Swedish Governmental Agency for Innovation Systems (VINNOVA) within the VINN Excellence center Chase. TG was supported by the US NSF under grant CCF-1535902 and by the US ONR under grant N00014-15-1-2676. CS was supported by Xilinx Inc., and by the US NSF under grants ECCS-1408006 and CCF-1535897. }
\thanks{Parts of this paper have been submitted to a journal~\cite{jacobsson16c}. The present paper focuses on higher-order modulation schemes and channel training aspects.}
}

\maketitle


\vspace{-1cm}
\begin{abstract}
%
Massive multi-user (MU) multiple-input multiple-output~(MIMO) is widely believed to be a core technology for the upcoming fifth-generation (5G) wireless communication standards. 
The use of low-precision digital-to-analog converters~(DACs) in MU-MIMO base stations is of interest because it reduces the power consumption, system costs, and raw baseband data rates.
In this paper, we develop novel algorithms for downlink precoding in massive MU-MIMO systems with 1-bit DACs that support higher-order modulation schemes such as \mbox{8-PSK} or 16-QAM. Specifically, we present low-complexity nonlinear precoding algorithms that achieve low error rates when combined with blind or training-based channel-estimation algorithms at the user equipment.
These results are in stark contrast to linear-quantized precoding algorithms, which suffer from a high error floor if used with high-order modulation schemes and 1-bit DACs.
\end{abstract}


\section{Introduction}
Massive multi-user (MU) multiple-input multiple-output (MIMO) is a promising technology for fifth-generation (5G) wireless communication standards, which enables substantial improvements in spectral efficiency, energy efficiency, reliability, and coverage compared to traditional multi-antenna systems. These gains are a result of equipping the base station~(BS) with hundreds of antennas and serving tens of user equipments~(UEs) in the same time-frequency resource~\cite{rusek14a, larsson14a}.
Increasing the number of radio frequency~(RF) chains at the BS leads, however, to a significant growth in system costs and circuit power consumption. Therefore, a successful deployment of massive MU-MIMO  requires the use of low-cost and power-efficient hardware components at the~BS.
This paper considers the downlink of massive MU-MIMO systems. 
We assume that the BS is equipped with 1-bit digital-to-analog converters (DACs) and transmits data using higher-order modulation schemes (such as 8-PSK or 16-QAM) to multiple~UEs.

\subsection{Benefits of Quantized Massive MU-MIMO}
Data converters at the BS are among the most dominant sources of power consumption in a massive MU-MIMO BS.  
Traditional multi-antenna BSs deploy high-resolution DACs (e.g., 10-bit or more) at each RF port. 
However, for massive MU-MIMO systems with hundreds or thousands of antenna elements, this  approach would lead to excessively high power consumption and system costs. 
A natural solution is to reduce the DAC resolution until the power budget and costs fall within tolerable levels.

\subsection{Relevant Prior Results}
While the impact of low-precision analog-to-digital converters (ADCs) on the massive MU-MIMO uplink has been studied extensively 
\cite{risi14a, jacobsson15a,studer15a, li16b,mollen16c}, far less is known about the use of low-precision DACs in the massive MU-MIMO downlink.
Recent results in~\cite{mezghani09c,saxena16a,usman16a} show that \emph{linear-quantized} precoders, which  perform traditional linear precoding followed by quantization, enable reliable transmission for relatively large antenna arrays in the high signal-to-noise ratio (SNR) regime, even in systems that use 1-bit DACs.  
\emph{Nonlinear} precoding algorithms have been proposed only recently in \cite{jacobsson16c, jedda16a}. 
Such precoding algorithms significantly outperform linear-quantized methods in the case of 1-bit DACs by approximating the optimal  precoding problem (which is of combinatorial nature) using, for example, convex relaxation techniques. All these results, although encouraging, focus on low-order modulation schemes such as QPSK. It is therefore an open question whether higher-order modulation schemes, such as 16-QAM, can be transmitted reliably in massive MU-MIMO systems that use 1-bit DACs. 

\subsection{Contributions}
We develop novel nonlinear precoding algorithms that relax the optimal 1-bit precoding problem and compute accurate solutions at low complexity. Our precoding algorithms rely on semidefinite and convex relaxation techniques to enable low-complexity precoding, even for systems with hundreds of antenna elements. We also investigate training-based and blind estimation techniques of the channel gain at the UEs. This estimation step is crucial for higher-order (and nonconstant modulus) constellations, such as 16-QAM. We demonstrate that the proposed  nonlinear precoding and channel-gain estimation algorithms enable reliable transmission of higher-order modulation schemes, for moderately-sized antenna arrays.

\subsection{Notation}
Lowercase and uppercase boldface letters designate column vectors and matrices, respectively. 
For a matrix $\bA$, we denote its transpose and Hermitian transpose by $\bA^T$ and~$\bA^H$, respectively. 
The entry on the $k$th row and  $\ell$th column of $\bA$ is $[\matA]_{k,\ell}$. 
The $k$th entry of the vector $\veca$ is~$[\veca]_k$.
We use $\matA \succeq \matzero$ to indicate that~$\matA$ is positive semidefinite.
%
%
The $M\times M$ identity matrix is denoted by $\bI_M$.
The real and imaginary parts of a complex vector $\veca$ are $\Re\{\veca\}$ and $\Im\{\veca\}$, respectively. 
We use $\sign(\cdot)$ to denote the signum function, which is applied entry-wise to a vector and defined as $\sign(a) = +1$ for $a \ge  0$ and $\sign(a) = −1$ for $a < 0$.
The $\ell_2$-norm and the~$\ell_\infty$-norm of $\veca$ are~$\vecnorm{\veca}_2$ and $\vecnorm{\veca}_\infty$, respectively; $\|\bA\|_F$ is the Frobenius norm of $\bA$.

\section{1-Bit Quantized Precoding}

\subsection{Downlink System Model}
\setlength{\textfloatsep}{15pt}
\begin{figure}[t]
\centering
 \includegraphics[width=0.99\columnwidth]{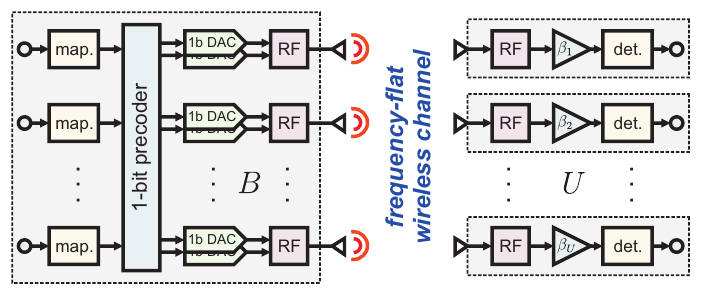}
 \caption{Overview of a massive MU-MIMO downlink system with 1-bit DACs. Left: a  massive MU-MIMO BS with $B$ antennas that performs 1-bit precoding to enable the use of 1-bit DACs; Right: $U$ single-antenna UEs. } 
  \label{fig:overview}
\end{figure}

As illustrated in \fref{fig:overview}, we consider a 1-bit massive MU-MIMO downlink system consisting  of a BS  with $B$ antennas that serves $U$ single-antenna UEs simultaneously and in the same frequency band.
We consider a block-fading scenario with the following narrowband input-output relation:
\begin{align} \label{eq:complex_channel}
\bmy[k] = \bH \bmx[k] + \bmn[k], \quad k=1,2,\ldots,K.
\end{align}
%
%
The vector $\vecy[k] = [y_1[k],\,\dots,\,y_U[k]]^T$ contains the received signals at all UEs, where $y_u[k] \in \complexset$ is the signal received at the $u$th UE in time slot $k$. 
The matrix $\matH  \in \opC^{U \times B}$ models the downlink channel, which is assumed to remain constant for~$K$ time slots and to be known perfectly at the BS. 
The vector $\bmn[k]\in\complexset^U$ in~\eqref{eq:complex_channel} models additive noise, which is assumed to be i.i.d.\ circularly-symmetric complex Gaussian with variance~$N_0$ per complex entry; the noise variance is assumed to be known perfectly at the~BS. 
The 1-bit precoded vector at time slot $k$ is denoted by $\bmx[k]\in\setX^B$, where~$\setX=\{\pm\ell\pm j\ell\}$ for a given (and fixed) \mbox{$\ell>0$} that determines the transmit power.
In what follows, we will often rewrite~\fref{eq:complex_channel} in the following equivalent matrix form: $\bY = \bH \bX + \bN$, with $\bY=\big[\bmy[1],\ldots,\bmy[K]\big]$,  $\bX=\big[\bmx[1],\ldots,\bmx[K]\big]$, and  $\bN=\big[\bmn[1],\ldots,\bmn[K]\big]$.

\subsection{Precoding in a Nutshell}
The goal of precoding is to transmit constellation points $s_u[k] \in \setO$ for $u=1,\ldots,U$ to each UE~$u$ at time slot $k$. Here,~$\setO$ is the constellation set (e.g., QPSK or 16-QAM).
The BS uses the knowledge of~$\matH$ to precode the symbol vector $\vecs[k] = \big[s_1[k],\,\dots,\,s_U[k]\big]^T$ into a $B$-dimensional precoded vector $\vecx[k] = \precode(\vecs[k], \matH)$. 
The function $\precode(\cdot, \cdot): \setO^U \times \opC^{U \times B} \rightarrow \setX^B$ represents the precoder.  
We assume that the precoding vectors $\vecx[k]$, $k=1,2,\ldots,K$, satisfy the instantaneous power constraint $\vecnorm{\vecx[k]}^2_2 = P$, and we define $\snr=P/\No$ as the SNR. 
For 1-bit DACs, this assumption leads to $\setX = \big\{\pm \ell \pm j \ell\big\}$ with $\ell=\sqrt{P/(2B)}$.

Coherent transmission of data using multiple BS antennas leads to an \emph{array gain} that depends on the channel matrix. We assume that the $u$th UE is able to rescale the received signals~$y_u[k]$, $k=1,2,\ldots,K$, by the so-called \emph{precoding factor} $\beta_u \in \opR^+$ in order to compute estimates~$\hat{s}_u[k]\in\complexset$ of the transmit symbol $s_u[k]\in\setO$ as~follows:
\begin{IEEEeqnarray}{rCl} \label{eq:scaling}
\hat{s}_u[k] = \beta_u y_u[k], \quad k=1,2,\ldots,K.
\end{IEEEeqnarray}
In \fref{sec:precodingfactorestimation}, we will discuss methods that enable the UEs to estimate the precoding factors $\beta_u$ for block-fading channels. 

In essence, the goal of precoding is to increase the signal power to the intended UEs while simultaneously reducing multi-user interference (MUI)~\cite{bjornson14b}. While there exist multiple formulations of this optimization problem based on different performance metrics, e.g., sum-rate throughput, worst-case throughput, or error probability (see~\cite{bjornson13a} for a survey), we focus exclusively on precoders that minimize the mean-square error (MSE) between the estimated symbol vectors~$\hat{\vecs}[k] = \big[\hat{s}_1[k],\ldots,\hat{s}_U[k]\big]^T$ and the transmitted symbol vectors $\vecs[k]$ under an instantaneous power constraint.

\subsection{Linear and Nonlinear Quantized Precoding}
\label{sec:quantprecoders}
In the infinite-resolution case, linear precoders multiply the symbol vector $\vecs[k]$ with a precoding matrix $\matP \in \opC^{B \times U}$ so that $\vecx[k]=\matP\vecs[k]$. This approach requires low  complexity and simple linear precoders, such as maximum ratio transmission (MRT) or zero-forcing (ZF), approach optimal performance in the large-antenna limit~\cite{rusek14a}. 
In the 1-bit case, linear-quantized precoders  perform linear precoding followed by quantization to the finite transmit set $\setX^B$ as
\begin{IEEEeqnarray}{rCl}
\vecx[k]  = \sqrt{\frac{P}{2B}}\big(\sign\lefto(\Re\lefto\{\matP\vecs[k]\right\}\right) + j\sign\lefto(\Im\lefto\{\matP\vecs[k]\right\}\right)\!\big)	
\end{IEEEeqnarray}
for each time slot $k=1,2,\ldots,K$.
%
%
Linear-quantized precoders have low complexity and their performance can be characterized analytically. 
However, nonlinear precoders significantly outperform such precoders~\cite{jacobsson16c}. 
The nonlinear precoders for block-fading systems considered in this paper minimize the total MSE  between all transmit symbols $\bS = \big[\bms[1],\ldots,\bms[K]\big]$ and their estimates $\widehat\bS = \beta\bY$ (over all $K$ time slots):
\begin{IEEEeqnarray}{rCl} \label{eq:mse}
\Ex{\bN}{\big\| \bS - \widehat\bS\big\|_F^2}
= \vecnorm{\bS - \beta\matH\bX}_F^2 + \beta^2UK \No.
\end{IEEEeqnarray}
Here, we have restricted ourselves to the case in which the precoder results in the same gain $\beta \in \opR^+$ for all UEs over all $K$ time slots. 
This expression allows us to formulate the MSE-optimal 1-bit quantized precoding~($\OBQP$) problem as
\begin{align} \label{eq:problem_complex_1bit}
\text{(\OBQP)} \quad 
\underset{\bX \in \setX^{B\times K},\, \beta \in \reals^+}{\text{minimize}}  \vecnorm{\bS - \beta \matH\bX}^2_F + \beta^2 U K \No
\end{align}
which simultaneously finds the optimal precoding vectors $\vecx^\text{QP}[k]$, $k=1,\ldots,K$,  and the associated precoding factor~$\beta^\text{QP}$. 
We emphasize that for fixed~$\beta$, the problem~($\OBQP$) is a closest vector problem that is known to be NP-hard; this implies that there is no known efficient algorithm. 
To enable near-optimal nonlinear precoding in practice, we will introduce in \fref{sec:nonlinearprecoders} approximate algorithms whose complexity is moderate even for large BS antenna arrays.

\section{Nonlinear Precoding for 1-Bit DACs}
\label{sec:nonlinearprecoders}

Since optimal 1-bit precoding is NP-hard, the use of a brute-force search would result in prohibitive complexity in massive MU-MIMO systems with hundreds of  BS antennas. We next propose two nonlinear precoding algorithms that yield accurate but approximate solutions at low computational complexity. 

\subsection{Problem Transformation}

Before detailing our algorithms, we rewrite the problem in~\eqref{eq:problem_complex_1bit} in a more convenient way.
We start by defining an auxiliary vector~$\bmb[k]=\beta\bmx[k]$. We further define $\matB = \big[\bmb[1],\ldots,\bmb[K]\big]$ and rewrite~\eqref{eq:problem_complex_1bit} in the following equivalent form:
\begin{align} \label{eq:problem_tweak_complex}
\underset{\matB \in \setB^{B \times K}}{\text{minimize}} \quad \vecnorm{\bS - \matH \bB}_F^2 + \frac{U N_0}{P} \vecnorm{\bB}^2_F
\end{align}
where $\setB = \big\{ \sqrt{P/(2B)}\,\lefto(\pm \beta \pm j\beta\right), \text{ for all } \beta > 0 \big\}$. Here, we have used the fact that $\beta^2 = \vecnorm{\matB}_F^2/(KP)$. 
Let $\bB^\text{QP}$ be the solution to~\eqref{eq:problem_tweak_complex}. Then, the resulting precoding vectors are obtained by scaling each entry of $\bB^\text{QP}$ so that it belongs to the set $\setX$ of 1-bit quantization outcomes.  

It will be convenient to transform the complex-valued problem~\eqref{eq:problem_tweak_complex} into an equivalent, real-valued problem using 
\begin{align*}
\bB_\opR =
\begin{bmatrix}
\Re\{\bB\} \\ \Im\{\bB\}
\end{bmatrix}\!,
\
\bS_\opR =
\begin{bmatrix}
\Re\{\bS\} \\ \Im\{\bS\}
\end{bmatrix}\!, \
\matH_\opR = 
\begin{bmatrix}
\Re\{\matH\} & -\Im\{\matH\} \\ 
\Im\{\matH\} & \Re\{\matH\}
\end{bmatrix}\!.
\end{align*}
These definitions enable us to rewrite \eqref{eq:problem_tweak_complex} as 
\begin{align} \label{eq:problem_tweak_real}
\underset{\bB_\opR \in \setB^{2B\times K}_\opR}{\text{minimize}} \quad \vecnorm{\bS_\opR - \matH_\opR \bB_\opR}_F^2 + \frac{U N_0}{P} \vecnorm{\bB_\opR}^2_F
\end{align}
where $\setB_\opR = \big\{ \pm \sqrt{P/(2B)}\,\beta, \text{ for all } \beta > 0 \big\}$.

As a last step, we vectorize the problem in \fref{eq:problem_tweak_real}. We make use of the vectorization operator $\text{vec}(\cdot)$ and the well-known  Kronecker product property $\text{vec}(\bA\bB\bC)=(\bC^T\kron\bA)\text{vec}(\bB)$. Since $\|\bA\|_F=\|\text{vec}(\bA)\|_2$, we can rewrite \eqref{eq:problem_tweak_real} as
\begin{align} \label{eq:problem_tweak_real1}
\underset{\bar{\vecb}_\opR \in \setB^{2B K}_\opR}{\text{minimize}} \quad \vecnorm{\bar\bms_\opR - \bar\matH_\opR \bar\bmb_\opR}_2^2 + \frac{U N_0}{P} \vecnorm{\bar\bmb_\opR}^2_2
\end{align}
where $\bar\vecb_\opR=\text{vec}(\bB_\opR)$ and $\bar\bH_\opR=\bI_{K}\kron\bH_\opR$.
We are now ready to detail our nonlinear precoding algorithms.

\subsection{Semidefinite Relaxation}

Semidefinite relaxation (SDR) is a well-established technique to approximately solve  a variety of discrete programming problems~\cite{luo10a}.  
In our case, proceeding  as in~\cite{jacobsson16c}, we relax~\eqref{eq:problem_tweak_real1} to the following semidefinite program (SDP): 
\begin{align*}
(\SDR) \, \left\{
 \begin{array}{ll}
\underset{\bar\matB_\opR \in \opS^{2BK+1}}{\text{minimize}} & \!\!\!\! \tr\lefto( \bar\matT_\opR \bar\matB_\opR \right) \\
\text{subject to} & \!\!\!\!  [\bar\matB_\opR]_{1,1} = [\bar\matB_\opR]_{b,b},  b=2,\ldots,2BK \\
& \!\!\!\! [\bar\matB_\opR]_{2BK+1, \, 2BK+1} = 1  \\
& \!\!\!\! \bar\matB_\opR \succeq \matzero.
\end{array}
\right.
\end{align*}
Here,  $\bar\matB_\opR = [\bar\vecb_\opR^T \ 1]^T [\bar\vecb_\opR^T \ 1]$ and 
\begin{IEEEeqnarray}{rCl}
\bar\matT_\opR &=& 
\begin{bmatrix}
\bar\matH_\opR^T\bar\matH_\opR + \frac{UN_0}{P}\matI_{2BK}& -\bar\matH_\opR^T\bar\vecs_\opR \\ 
-\bar\vecs_\opR^T\bar\matH_\opR & \vecnorm{\bar\vecs_\opR}^2_2
\end{bmatrix}\!.
\end{IEEEeqnarray}
We note that the key difference between $(\SDR)$ and the precoding problem given in~\cite{jacobsson16c} is that the method presented here is for the block fading channel in~\eqref{eq:complex_channel} with transmission over~$K$ time slots; the method in~\cite{jacobsson16c} considers a single time-slot only, i.e., it deals with the special case of $K=1$.

If the solution $\bar\bB_\opR^\SDR$ has rank one, then SDR found the exact solution to the 1-bit precoding problem in~\fref{eq:problem_tweak_real1}. If, however, the rank of  $\bar\bB_\opR^\SDR$ exceeds one, then we have to extract a precoding vector that belongs to the discrete set~$\setX^{BK}$.
Such a vector can be obtained by first performing an eigenvalue-decomposition of~$\matB_\opR^\SDR$ followed by quantizing the first $2BK$ entries of the leading eigenvector (see~\cite{jacobsson16c} for the details).

The problem $(\SDR)$ can be solved via standard convex optimization methods, whose  worst-case complexity scales as  $(BK)^{4.5}$~\cite{luo10a}. 
Unfortunately, SDR lifts the problem to a higher dimension: from~$2BK$ dimensions to $(2BK+1)^2$ dimensions. Hence, even for a small number of time slots and/or BS antennas, the memory requirements and computational complexity of this approach becomes prohibitively large. Furthermore, implementing numerical solvers for SDP entails, in general, high hardware complexity~\cite{castaneda16a}. 
Hence, for large antenna arrays and a large number of time slots, alternative precoding algorithms are necessary. A suitable method that avoids lifting the problem to a higher dimension and requires low computational complexity is described next.

\subsection{Squared $\ell_\infty$-Norm Relaxation}
We start by rewriting the real-valued problem in~\eqref{eq:problem_tweak_real1}~as 
\begin{align} \label{eq:ett_problem}
\begin{array}{ll}
\underset{\bar\vecb_\opR \in \opR^{2BK}}{\text{minimize}} & \vecnorm{\bar\vecs_\opR - \bar\matH_\opR \bar\vecb_\opR}_2^2 + \dfrac{2 UBK N_0}{P} \vecnorm{\bar\vecb_\opR}^2_\infty \\
\text{subject to} & [\bar\bmb_\opR]_1^2 = [\bar\bmb_\opR]_b^2 \text{ for } b = 2,\dots,\,2BK.
\end{array}
\end{align}
Here, we used that $\big\|{\bar\vecb_\opR}\big\|_2^2 = 2BK\big\|{\bar\vecb_\opR}\big\|_{\infty}^2$. By dropping the nonconvex constraints $[\bar\bmb_\opR]_1^2 = [\bar\bmb_\opR]_b^2$ for $b = 2,\,\dots,\,2BK$, we obtain the following convex relaxation of \eqref{eq:ett_problem}
\begin{align*}  
(\LINF)\, \,\, \underset{\bar\vecb_\opR\in \opR^{2BK}}{\text{minimize}}\, \, \vecnorm{\bar\vecs_\opR - \bar\matH_\opR \bar\vecb_\opR}_2^2 + \frac{2 U B K N_0}{P} \vecnorm{\bar\vecb_\opR}_{\infty}^2
\end{align*}
which we can solve efficiently using the squared $\ell_\infty$-norm relaxation algorithm (SQUID, for short) proposed in~\cite{jacobsson16c}. Each iteration of the SQUID algorithm requires only simple matrix-vector operations.
Hence, SQUID enables nonlinear precoding for very large antenna arrays and large number of time slots. 

 
\section{Estimating the Precoding Factor $\beta$}
\label{sec:precodingfactorestimation}
Accurate estimates of the precoding factor $\beta$ are crucial when one uses higher-order constellations  that are not of constant modulus, such as 16-QAM.
We next discuss two methods that enable each UE to acquire an accurate estimate of $\beta$.

\subsection{Pilot-Based Estimation}
A straightforward way to acquire an estimate $\hat\beta_u$ of the precoding factor $\beta$ at the $u$th UE is to use pilots that are known at the UE side.
We propose to transmit a pilot signal in the first time slot ($k=1$), i.e., we set $s_u[1]=\sqrt{E_s}$ for all $u=1,2,\ldots,U$. The remaining $K-1$ time slots can then be used for payload transmission. 
By transmitting the precoding vector~$\bmx[k]$ obtained from \fref{eq:problem_complex_1bit}, the effective input-output relation for the $u$th UE is given by 
\begin{align} \label{eq:effectivesystem}
y_u[k]= \beta^{-1}s_u[k]+ e_u[k] +n_u[k]
\end{align}
where $e_u[k]$ contains quantization errors and residual MUI. Assuming that $e_u[k] + n_u[k]$ is Gaussian distributed and independent of $s_u[k]$, each UE can compute a maximum-likelihood estimate (MLE) for $\beta$ as follows:
\begin{align*}
\hat\beta_u^\text{MLE} = \Re\big\{\sqrt{E_s}/y_u[1]\big\}, \quad u=1,2,\ldots,U.
\end{align*}
While one could transmit a large number of pilot symbols to enable more accurate estimates of the precoding factor, our results in \fref{sec:numericalresults} reveal that one pilot signal is typically sufficient to enable reliable downlink communication. 
 
\subsection{Blind Estimation}
An alternative method to acquire an  estimate of $\beta_u$ for the $u$th UE is to use blind estimation. 
The advantage of this approach is that all time slots can be used for data transmission. 
Assume that the transmit signals $s_u[k]$, residual  errors~$e_u[k]$, and the noise $n_u[k]$ are zero mean and independent. Then, the sample variance of the received signals at the $u$th UE satisfies $\frac{1}{K}\sum_{k=1}^K |y_u[k]|^2 \to \Ex{}{|y_u|^2}$ as $K\to\infty$.
From~\fref{eq:effectivesystem}, it follows that the variance is given by 
\begin{align*} 
\Ex{}{|y_u|^2} & = \Ex{}{|\beta^{-1}s_u + e_u+n_u|^2}  = \beta^{-2}E_s + E_0 + \No
\end{align*}
where $E_s = \Ex{}{|s_u|^2}$ is the average symbol energy, $E_0=\Ex{}{|e_u|^2}$ is the average error energy, and $\No=\Ex{}{|n_u|}^2$ is the noise variance. By noting that $\frac{1}{K}\sum_{k=1}^K |y_u[k]| \approx \Ex{}{|y_u|^2}$ for sufficiently large $K$ and by assuming that $E_s$, $E_0$, and $\No$ are known at the UE side, we propose that each UE computes a blind estimate for $\beta$ as follows:
 \begin{align*}
\hat\beta_u^\text{blind} = \sqrt{\frac{E_s}{\frac{1}{K}\sum_{k=1}^K|y_u[k]|^2-E_0-\No}}, \quad u=1,2,\ldots,U.
\end{align*}
While $E_s$ and $\No$ are typically known at the UE, $E_0$ is generally unknown as it depends on the precoding algorithm, the channel statistics, and the transmit symbols. 
We therefore set $E_0=0$, which---as we will shown in \fref{sec:numericalresults}---yields sufficiently accurate estimates and enables reliable transmission, even for a small number of time slots~$K$.\footnote{Hybrid methods that combine pilot-based and blind estimation techniques may further improve the accuracy of the estimation of $\beta$.}

\section{Numerical Results}
\label{sec:numericalresults}

\begin{figure}[t]
\centering
\subfloat[Linear-quantized ZF precoding.]{\includegraphics[width=0.95\columnwidth]{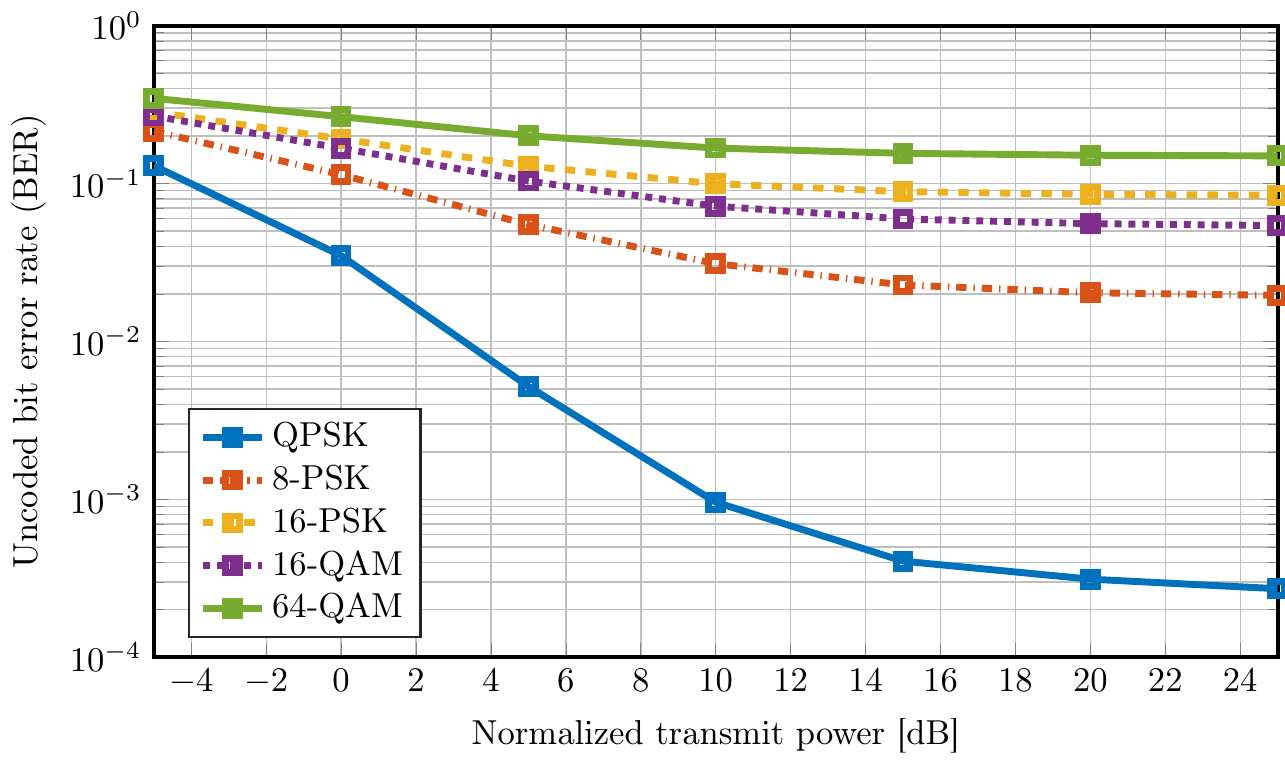}\label{fig:asilomar_modulation_zfq}}
\quad
\subfloat[Nonlinear precoding via SQUID.]{\includegraphics[width=0.95\columnwidth]{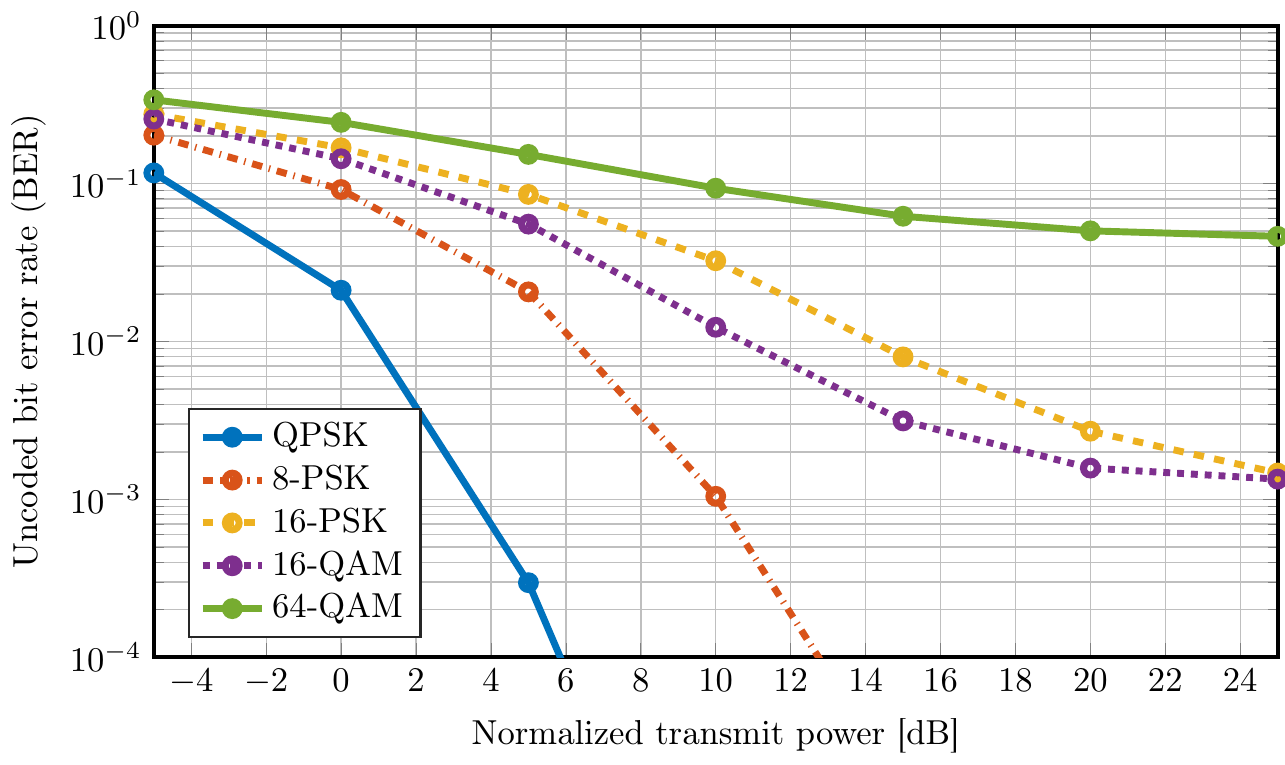}\label{fig:asilomar_modulation_squid}}
\caption{Comparison of different modulation schemes with linear-quantized ZF precoding (a) and nonlinear SQUID precoding (b) for a $128$-BS-antenna, $16$-UE massive MU-MIMO system with blind $\beta$ estimation ($K=10$). }
\label{fig:asilomar_modulation}
\end{figure}

We now present numerical simulation results for 1-bit precoding with higher-order modulation schemes. Throughout this section, we consider the bit-error rate (BER) for uncoded transmission as the main performance metric. 
Furthermore, we focus on  i.i.d.\ Rayleigh fading channel matrices. 

\subsection{Comparison of Modulation Schemes}
\fref{fig:asilomar_modulation} compares the performance of several constellations for the case of linear-quantized ZF precoding~\cite{jacobsson16c} and nonlinear SQUID-based precoding in a $128$-BS-antenna, $16$-UE massive MU-MIMO system with blind $\beta$ estimation over $K=10$ time slots. 
We make the following observations: (i) nonlinear precoding significantly outperforms linear-quantized precoding for all modulation schemes; (ii) nonlinear precoding enables reliable transmission of modulation schemes of higher order, such as 8-PSK, 16-QAM, and 16-PSK; (iii) 16-QAM outperforms 16-PSK transmission, which implies that tightly packing constellation points is advantageous even if this requires an accurate estimate of $\beta$ (note that for constant-modulus constellations, such as QPSK, 8-PSK, and 16-PSK, the precoding factor does not need to be estimated under minimum distance decoding~\cite{jacobsson16c}); and (iv) 64-QAM results in a high SNR floor and reliable transmission would require either more BS antennas or forward error correction.

\subsection{Pilot-Based vs. Blind $\beta$ Estimation}

\begin{figure}[t]
\centering
\includegraphics[width=0.95\columnwidth]{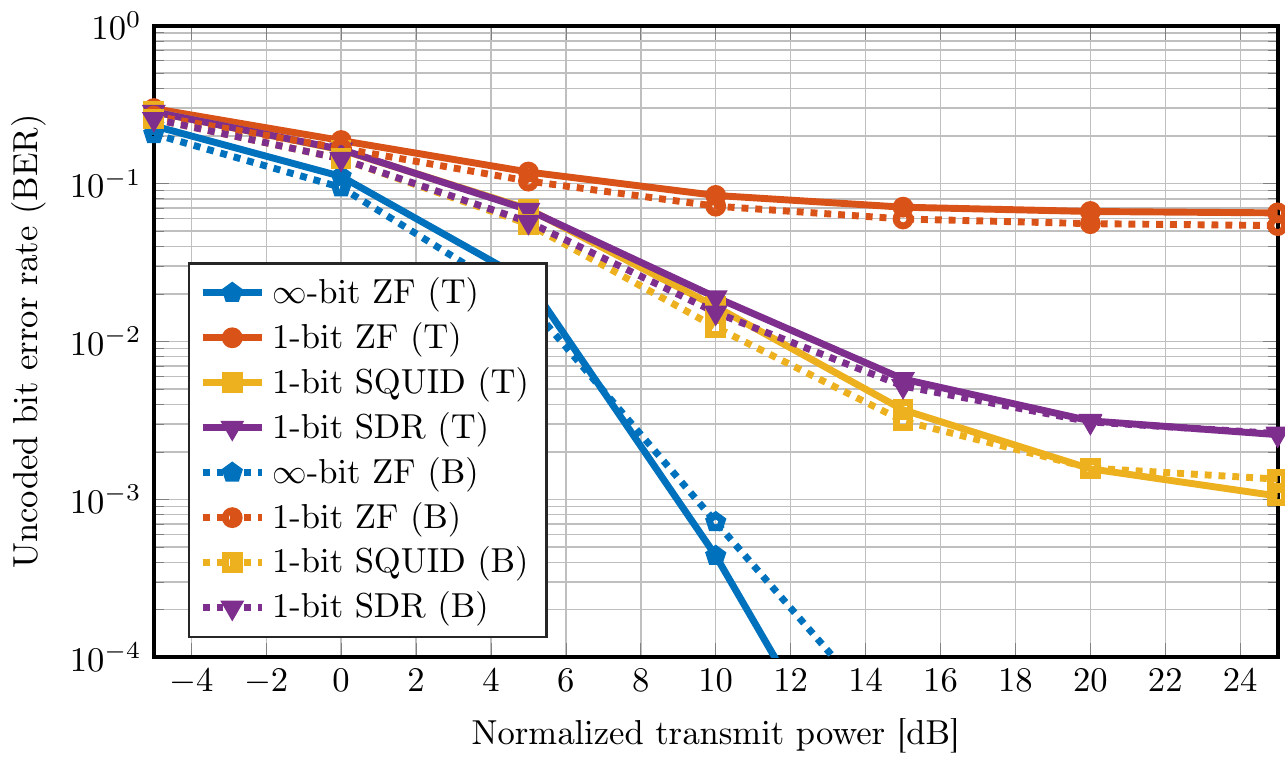}
\caption{Comparison between training-based (T) and blind (B) $\beta$ estimation; 16-QAM, $B = 128$~BS antennas, $U = 16$~UEs, and $K = 10$~time slots.}
\label{fig:asilomar_blindvstraining}
\end{figure}

\begin{figure}[t]
\centering
\includegraphics[width=0.95\columnwidth]{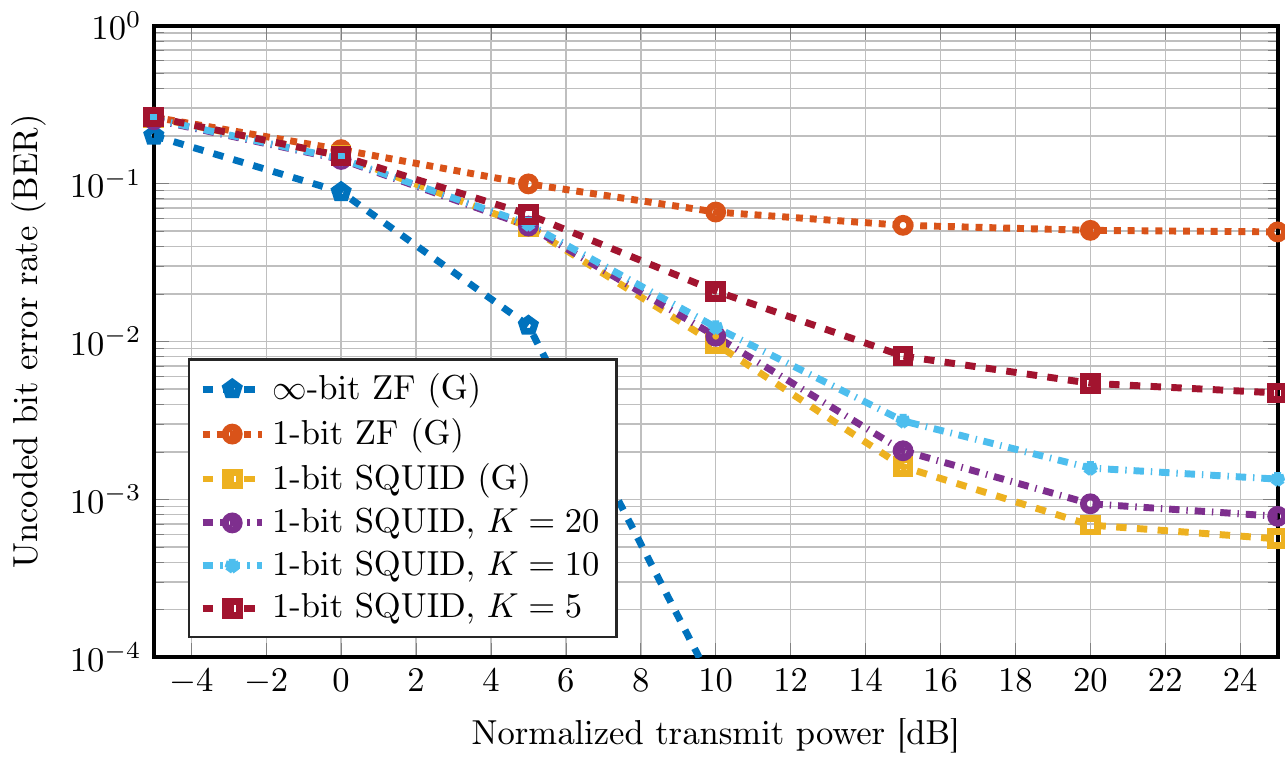}
\caption{Impact of the number of time slots $K$ for blind estimation; 16-QAM, $B = 128$~BS antennas, and $U = 16$~UEs.}
\label{fig:asilomar_training_sweep}
\end{figure}

\fref{fig:asilomar_blindvstraining} compares the BER with 16-QAM signaling for training-based and blind estimation of the precoding factor~$\beta$ in a $128$-BS-antenna, $16$-UE massive MU-MIMO system and $K=10$ time slots.  We see that both estimation methods yield similar BER performance, independent of the precoding algorithm. 
\fref{fig:asilomar_training_sweep} shows,  the impact of the number of time slots~$K$ over which the precoding factor~$\beta$ is computed via blind estimation. We see that only a few time slots (e.g., $K=10$) are sufficient to achieve near genie-aided~(G) performance (i.e., assuming perfect knowledge of $\beta$ at the UEs). We furthermore see that SDR-based precoding performs slightly worse than SQUID-based precoding. The reason is that our implementation of SDR-based precoding operates over a single time slot, to minimize the complexity; SQUID, in contrast, is able to solve the block-based nonlinear precoding problem.

\section{Conclusions}

We have investigated the performance of higher-order modulation schemes for massive MU-MIMO downlink systems with 1-bit ADCs.
We have developed two low-complexity, nonlinear precoding algorithms suitable for block-fading transmission and large BS antenna arrays. To enable reliable transmission of higher-order constellations, such as 16-QAM, we have proposed two algorithms for estimating the precoding factor $\beta$ at the UE side. Our simulation results demonstrate that (i) nonlinear precoding algorithms significantly outperform linear-quantized methods, (ii) higher-order modulation schemes, such as 8-PSK, 16-QAM, and 16-PSK can be transmitted reliably with nonlinear precoding algorithms, and (iii) 64-QAM requires large BS antenna arrays in combination with coding. Put simply, 1-bit massive MU-MIMO enables the use of low-cost and low-power RF circuitry, while supporting  high data rates. 

\bibliographystyle{IEEEtran}
\bibliography{IEEEabrv,confs-jrnls,publishers,studer,svenbib}

\end{document}